# Voltage Temperature Monitoring System (VTMS) for a BTS Room

Sadeque Reza Khan[1], Siddique Reza[2] and Arifa Ferdousi[3]

[1]Dept. of Electronic and Communication Engineering, NITK, India
sadeque_008@yahoo.com
[2]Dept. of Computer Science and Engineering, MIST, Dhaka University, Dhaka, Bangladesh
remond_007@yahoo.com
[3]Dept. of Computer Science and Engineering, Varendra University, Rajshahi, Bangladesh
arifaferdousi@yahoo.com

## ABSTRACT

*Although Cellular communication is getting more and more popular in our country present days, but its network improvement is hampered by the crysis of electricity. The recent decision of present Government is that they will not provide any electricity from the grid to any new BTS rooms of any Celluler operator companies like Grammen Phone, Robi, Airtel etc. These companies have to develop their own power stations either by using generators or by developing solar plants. Now a days most of the BTS rooms, that the cellular operators are installing with a generator and 48 volt battery backup. So for the synchronisation of the operation of PDB, Generator and battery, they require a device called Voltage Temperature Monitoring System or VTMS. It is a Microcontroller based controlling unit which controlls the operation of generator and battery when PDB in not available in the BTS room.*

# 1. INTRODUCTION

In the history of Bangladesh, power crisis has reached to the worst-ever level, especially during the hot summer days, when country's average temperature is 31 to 35 degree Celsius [1]. Bangladesh has been facing electricity shortage for many years. The distribution system in Bangladesh is facing tremendous pressure from industrial and residential users to maintain a regular supply [2]. In last few years this problem was not serious but in the year of 2010-11 the problem has exceeded the common people's patient. Power outage/failure is a common phenomenon now-a-days and people are facing severe electrical load shedding, voltage fluctuation throughout the day and this problem is more severe in the rural areas, although only 30% of the total population enjoys the electricity facilities [3]. One of the major steps that our Goverment has taken to solve this electricity problem is that they will not provide any power from the grid to any new BTS and for new BTS, operators must have to develop their own power system by using generators or renewable energy [4], i.e. solar energy [5]. But many important devices and chip components of a BTS demand for stable temperature and voltage without instantaneous breakdown and wide range of fluctuation [6]. Early detection of overheating and proper handling of such situation in a BTS room is really essential [7] to avoid deterioration and faulty components. To solve the power problem, now-a-days most of the BTS rooms are installed with a diesel generator [8] and 48 volt battery backup. So they alternate the source like they use generator for 6 hours and battery for next 6 hours to fullfil their power requrements with the help of a controlling device called VTMS. And temperature is also maintained in a certain level by tracking the temperature with VTMS which provides an alarm for an overheating environment of BTS. So Voltage and Temperature Monitoring System (VTMS) is a micro-processor based digital programmable device, designed with integrated systems [9] [10]. It is used to control the whole power distribution environment by monitoring 48 volt battery voltage level and temperature of a BTS room, synchronizing and maintaining the supply among PDB, Generator and 48 volt battery supply and also by generating some important alarm as per required.

## 2. PROPOSED SYSTEM

The system contains a temperature sensor, a multi turn 5 kilo ohm resistor to adjust the battery voltage level and some relays to generate different alarms and also to sense the mains fail and low fuel condition. The whole system is controlled by a PIC microcontroller 16F877A. It senses whether the PDB is available or not if not then takes different decisions, synchronize the operation of generator and battery and give some alarms as per logic provided. The overall system is shown in Fig 1.

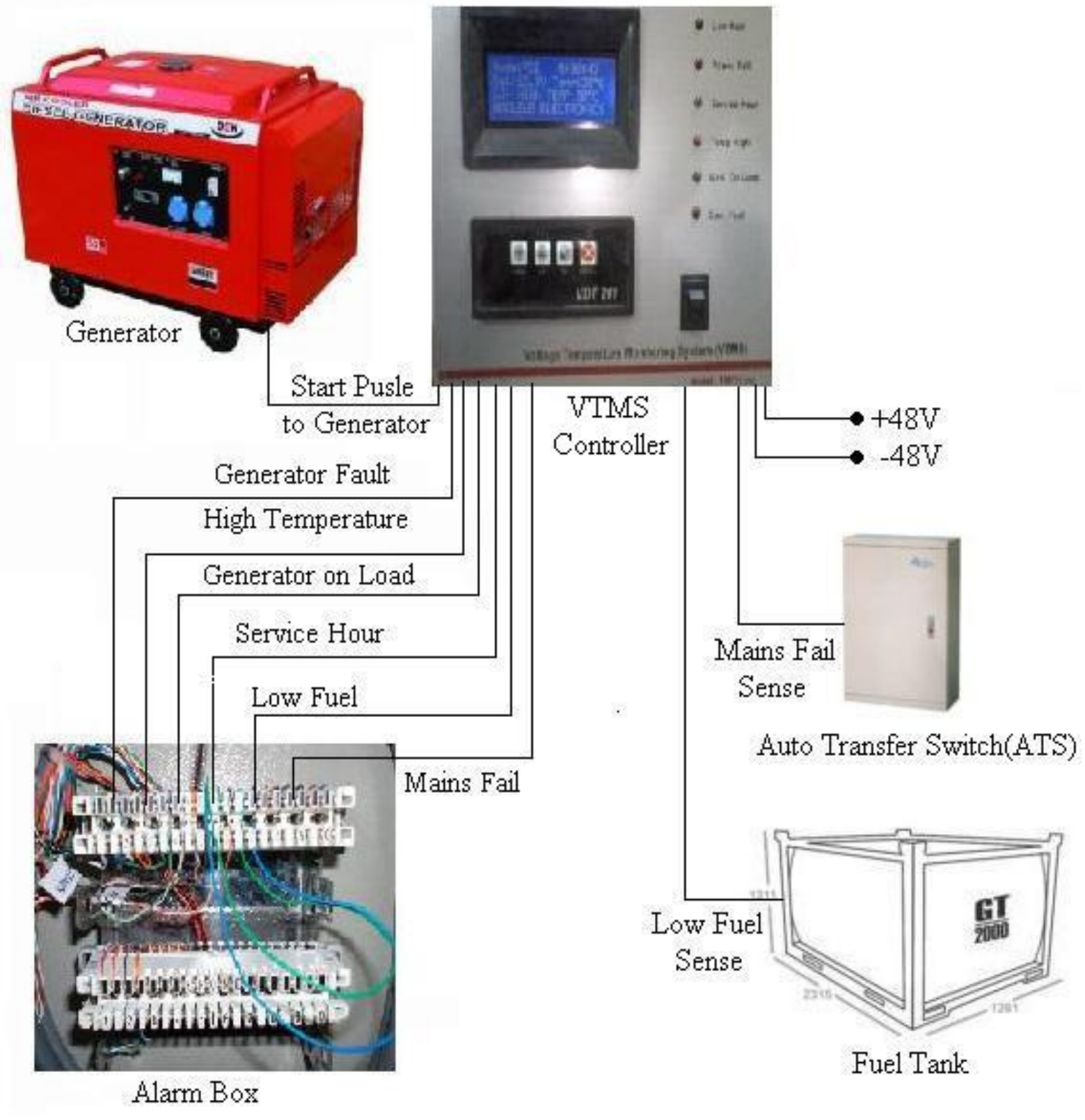

Figure 1. System overview

# 3. SYSTEM DESIGN

The design steps and working principles of the system is organized into two different units like Hardware unit and Software unit. Hardware unit includes controller unit, power supply section, reset section, display section, sensor unit, and battery voltage monitoring system and alarm section. Software unit includes the compiler to build the assembly program used in PIC microchip.

## 3.1 Hardware design

### 3.1.1 Controller unit

The control module is built with the microcontroller IC. The central controller is Microchip PIC16F877A. Microcontroller is one of the ways of the evolution of microprocessor. It consists of a microprocessor, RAM, EEPROM or EPROM, I/O capacities, ADC, timer, interrupt [11] controller and embedded controller. The microcontroller chip has the versatility to sense inputs and control outputs in the devices. PIC 16F877A is an upper range and 16 series low cost 8 bit microcontroller [12] [13]. It consists of 33 I/O (Bi directional lines) with 25mA current in per pin. It also has 5channel built-in A/D converter [14] and serial communication [15].

### 3.1.2 Power supply section

All the BTS rooms contain rectifier which provides 48 volt DC . So it is necessary to convert this 48 volt DC to 5 volt DC which is not generally possible wihout any SMPS(Switch Mode Power Supply) [16] [17] [18]. So a small power supply is developed for these BTS room products and it is common part for all controller of BTS. It can also be defined as a small version of SMPS.

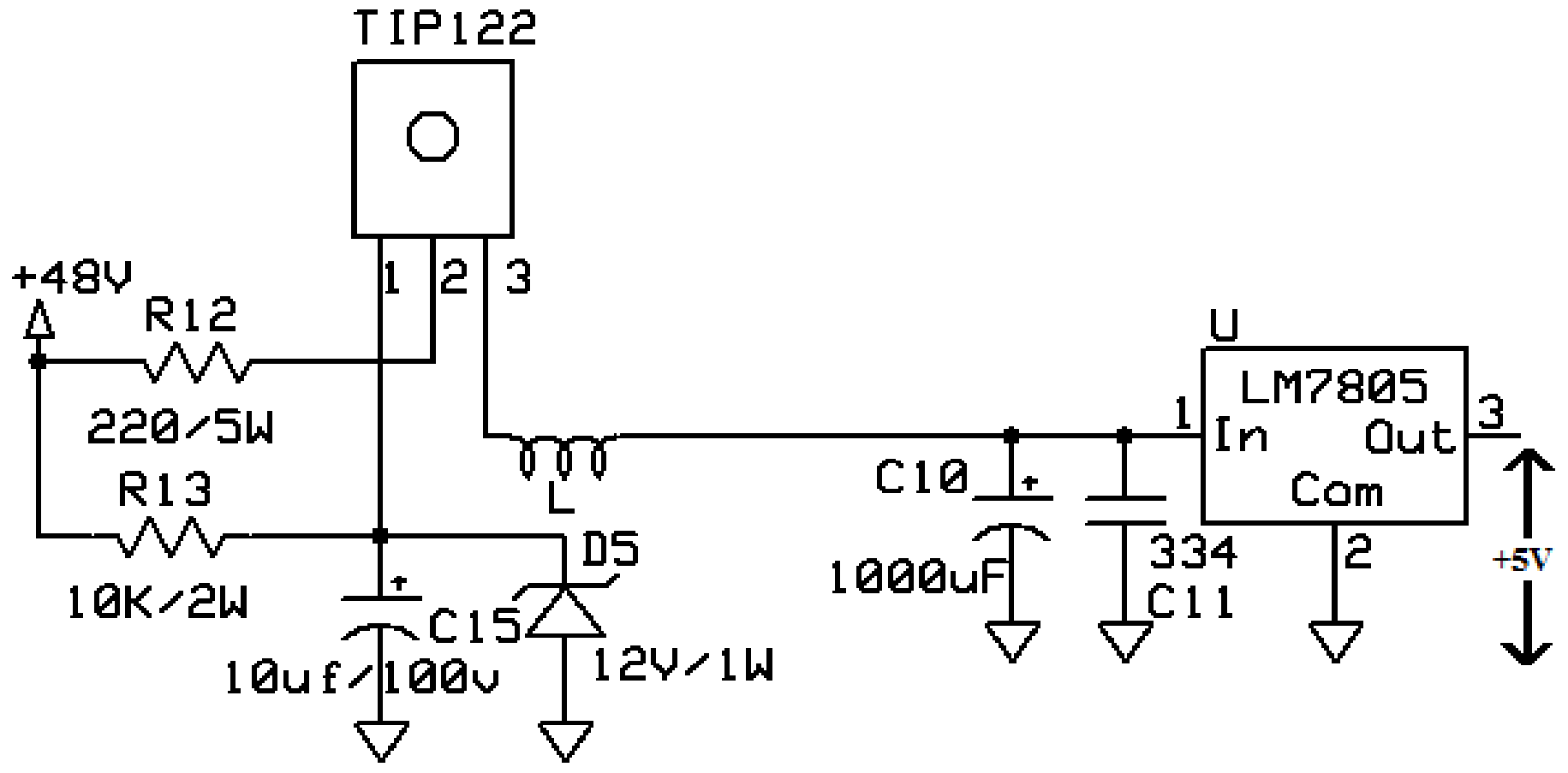

Figure 2. Power supply section of VTMS

### 3.1.3 Reset section

It is the most important part of this whole controller. Because it is working as an external watchdog timer [19] for the microcontroller. BTS room is full of thermal noise [20]. So the possility of getting hung is so much, that it is necessary to set up a portion in the main circuit to watch the microcontroller periodically. So, if the main controller gets hung then the reset circuit will recover it from this condition. The main work is done here by CD4047 IC [21] wihch is a CMOS monostable/astable multivibrator [22] [23]. This IC is edge triggerd which is the most significant quality of it.

### 3.1.4 Display section

For display section a 4x20 line LCD (Liquid Crystal Display) is used. LCD is now a very common choice for graphical and alphanumeric displays. Generally LCD is a high contrast control module with a 4-bit or 8-bit data bus and built in temperature control module [24] [25].

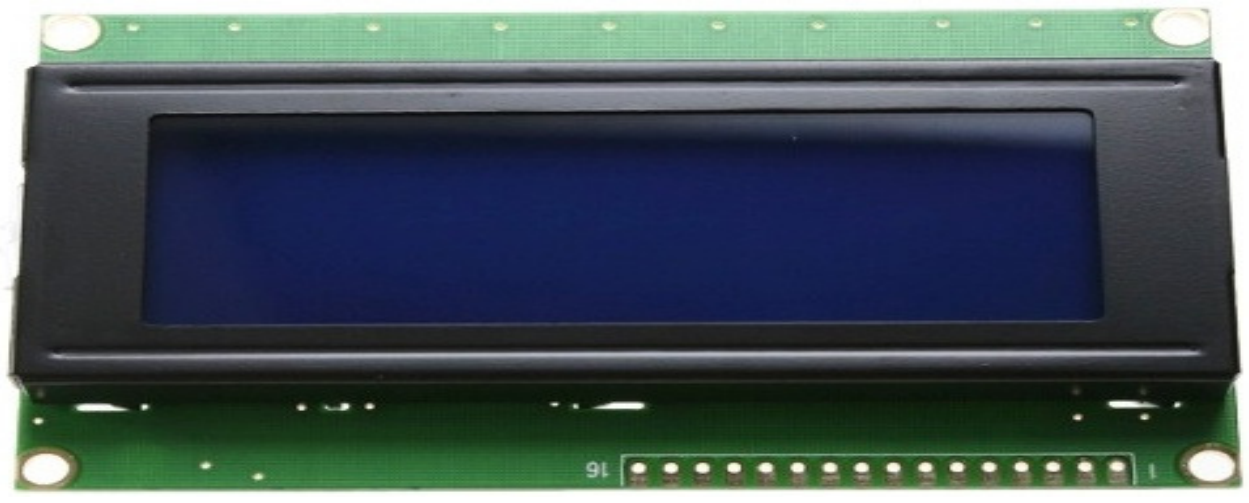
\

Figure 3. 4X20 Line LCD

### 3.1.5 Sensor section

LM35DZ is used here as a temperature sensor to measure the room temperature. It is a linear device or it provides linear data. The LM35 series are precision integrated-circuit temperature sensors, whose output voltage is linearly proportional to the Celsius (Centigrade) temperature [26] [27] [28]. The calculation of temperature from the sensor is shown in equation 1.

**Temperature = Voltage level in ADC0 (AN0)/2 …………………………………………..Eqn.1**

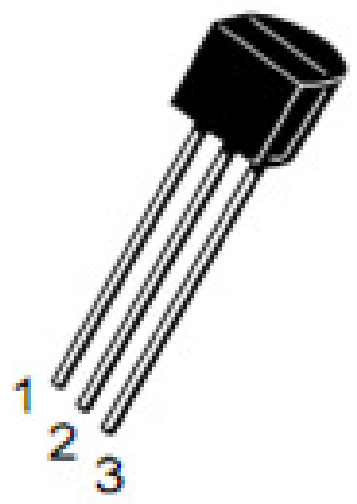

Figure 4(a). LM35DZ[5]

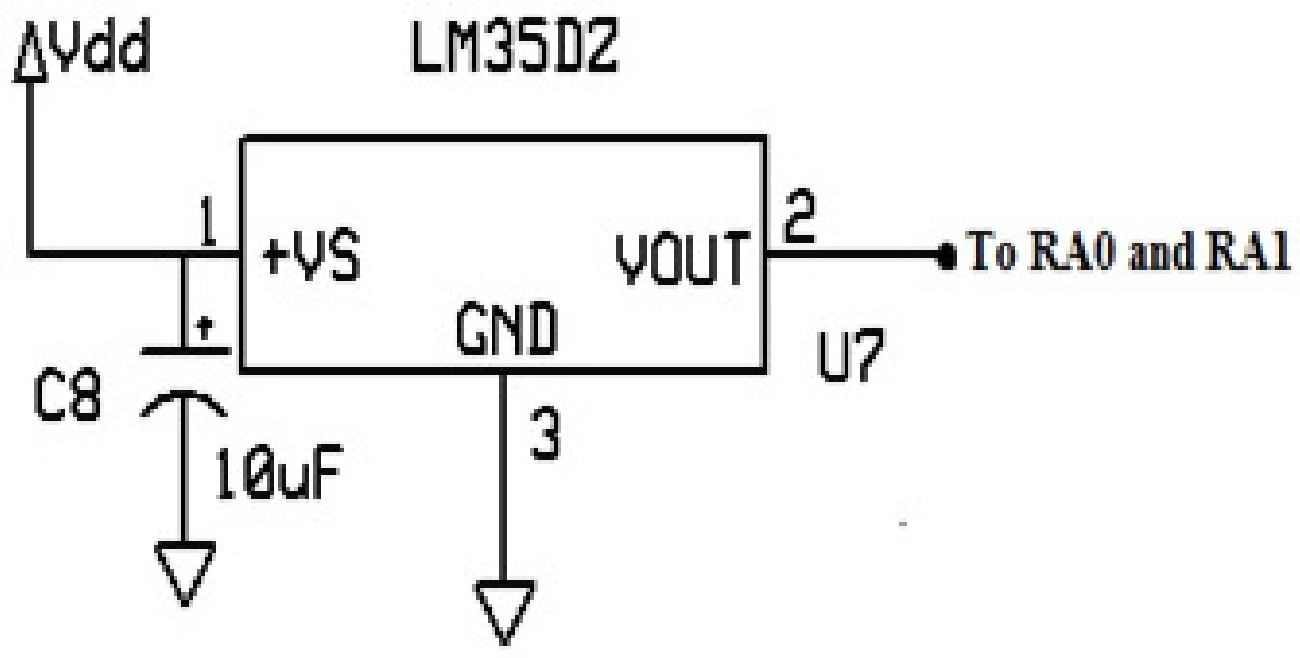

Figure 4(b). Connection Diagram of LM35DZ[5]

### 3.1.6 Battery voltage monitoring system

Battery voltage level monitoring [29] is one of the important part of this device. So we can monitor the battery voltage level of a BTS room by the given circuit. For the precise adjustments of voltage a 5kohm multiturn is used. Multiturn trimmers are suitable for applications that require fine resistance adjustment [30] [31]. Battery voltage and its decimal value calculation are provided in equation 2 and equation 3 and total battery voltage calculation is provided in equation 4.

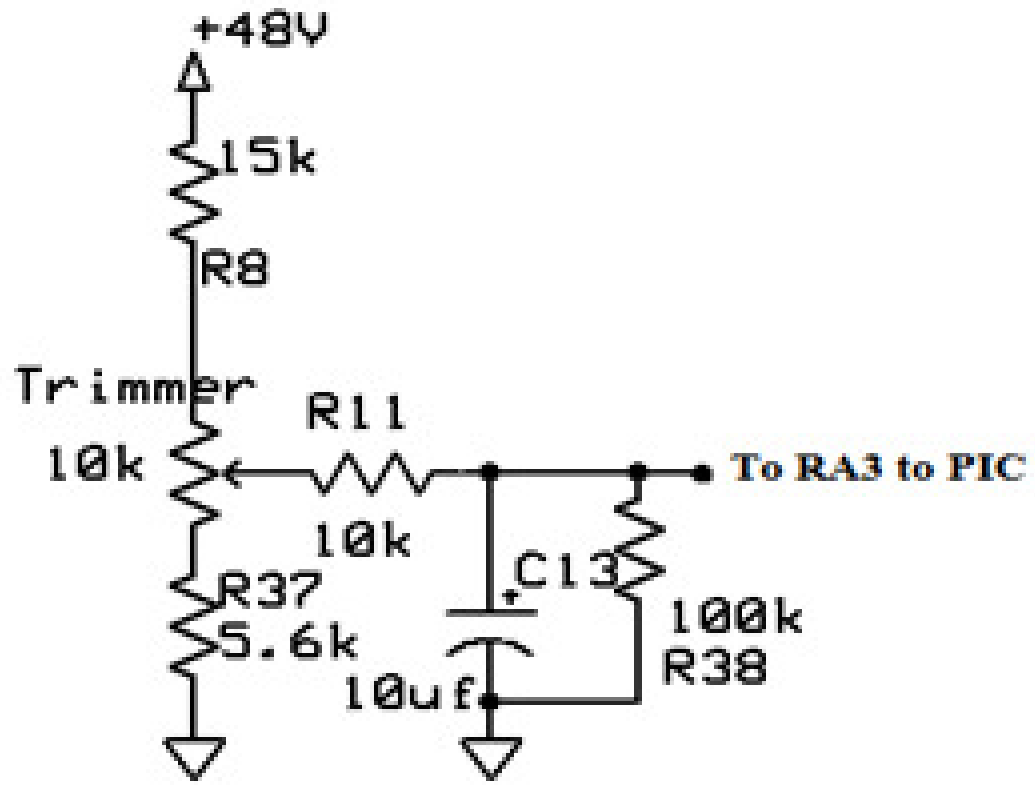

Figure 5(a). Battery Voltage Monitoring section

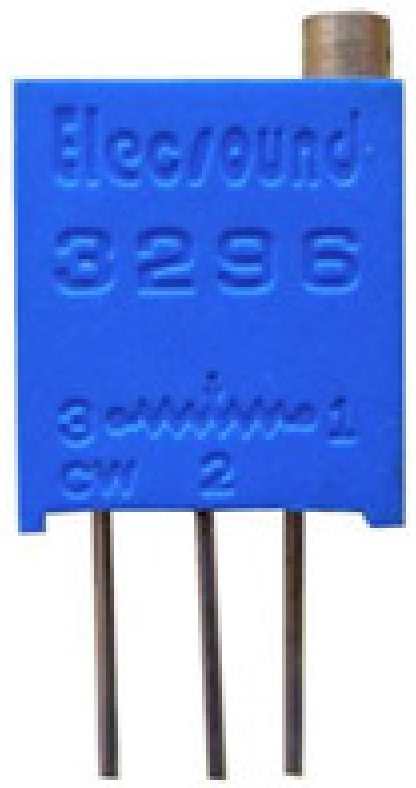

Figure 5(b). 5kohm multiturn

**Battery Voltage (Integer) = Voltage level in ADC1 (AN1)/16 ………………………………………… Eqn.2**

**Decimal value of Battery Voltage (Integer) = {(ADC1 MOD 16) * 10}/16 …………………………... Eqn.3**

**Total Battery voltage = Battery Voltage (Integer) + Decimal point (.)+ Decimal value (Integer) ….. Eqn.4**

### 3.1.7 Mains fail and Low fuel sense

For mains fail and low fuel sensing this controller uses two extra relays and also some security section which keeps the microcontroller away from noise and any other unwanted circumstance. Both those relays are controlled from the ATS (Auto Transfer Switch) and Fuel tank switch shown in Fig 6.

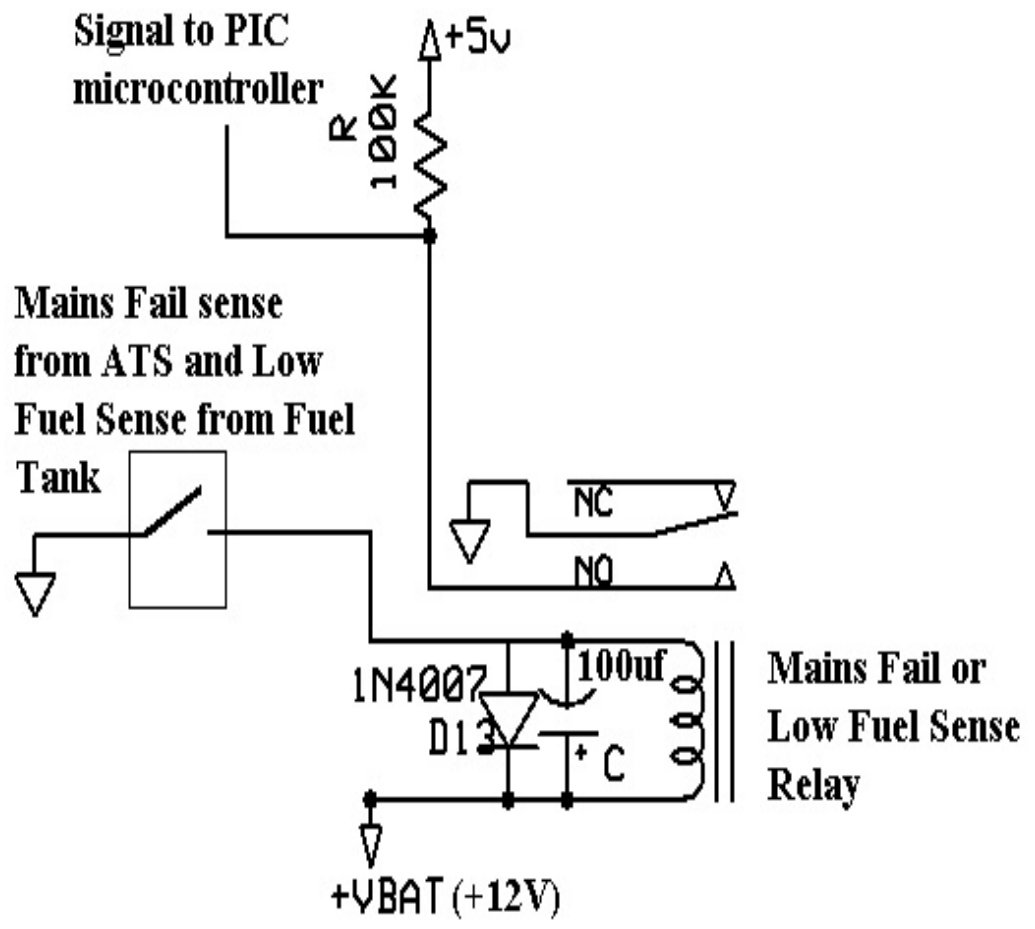

Figure 6. Mains fail and Low fuel sense

### 3.1.8 Auto Bypass

It is a special feature of VTMS. If because of any unexpected reasons the controller shuts down, it will automatically turn the generator on for the safety purpose of BTS room.

### 3.1.9 Alarm section:

This whole device is designed with six alarms. They are : (1) Mains fail, (2) Low fuel , (3) Genset on load, (4) High temerature, (5) Genset Fault and (6) Service hour alarm. These six alarms are given by the main controller to the alarm box of the BTS room through a relay contact which is shown in Fig 7.

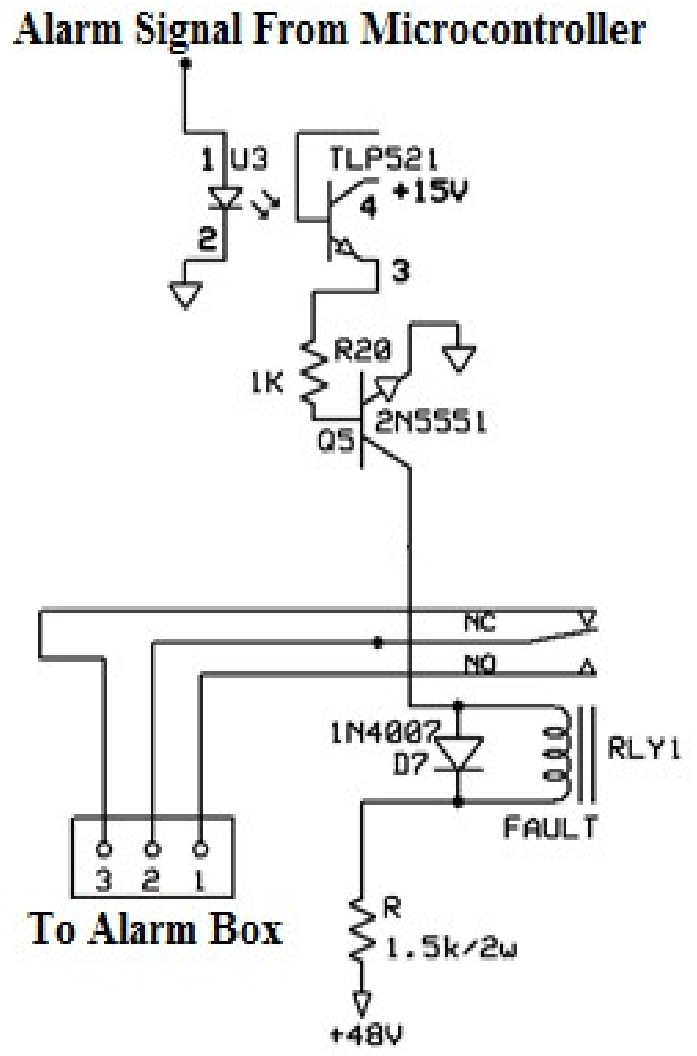

Figure 7. Alarm Section

## 3.2 Software design

### 3.2.1 MPLAB v 8.40

In this system MPLAB v8.40 is used to develop the program for PIC microchip. This compiler consists of Hitech C as well. So this compiler can be used to program in C language. Here the program is divided into fifteen macros. The main macro controls the whole program. There are eight macros to show some certain lines in the LCD. Two macros are used to measure the ADC value of temperature and battery voltage. Other three macros are designed to maintain the overall operation of the controller. Other three macros are used to display some special conditions like the setting option and the preset values.

### 3.2.2 Flow Chart

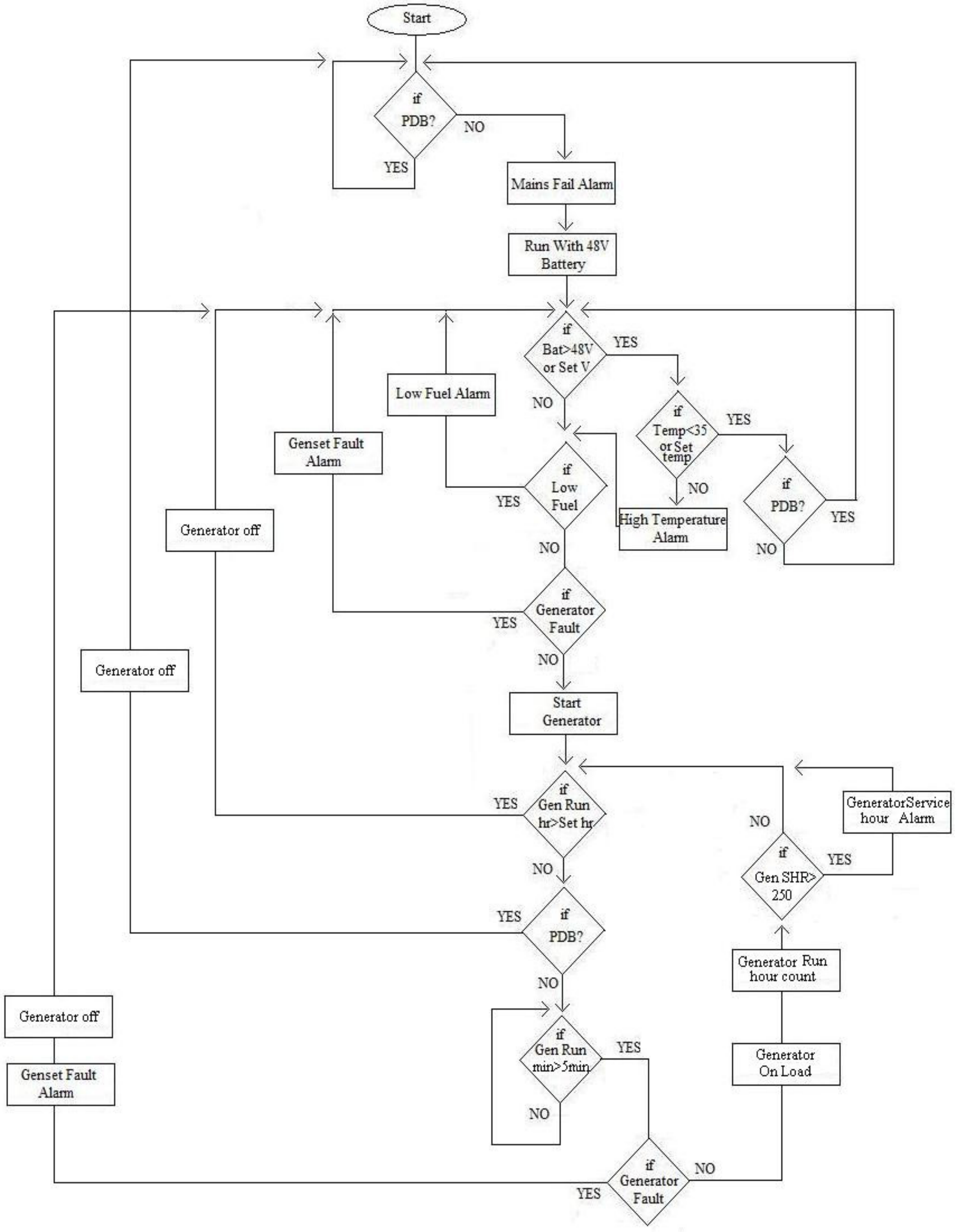

Flow chart shows whole logic of VTMS controller. This full logic is implemented in the main controller IC.

# 4. RESULT:

The result shows the complete controller in fig 8(a). This is the complete product and it is also tested in the real time environment. Fig 8(b) shows the ports to connect the wires coming from alarm box and also from generator. This figure shows the multiturn of battery voltage monitoring section is kept in such a way that battery voltage can be calibrated for showing in the LCD. This figure also shows the temperature sensor position. In fig 8(c) the LCD output is shown and in fig 8(d) the indicator LED and their names are shown.

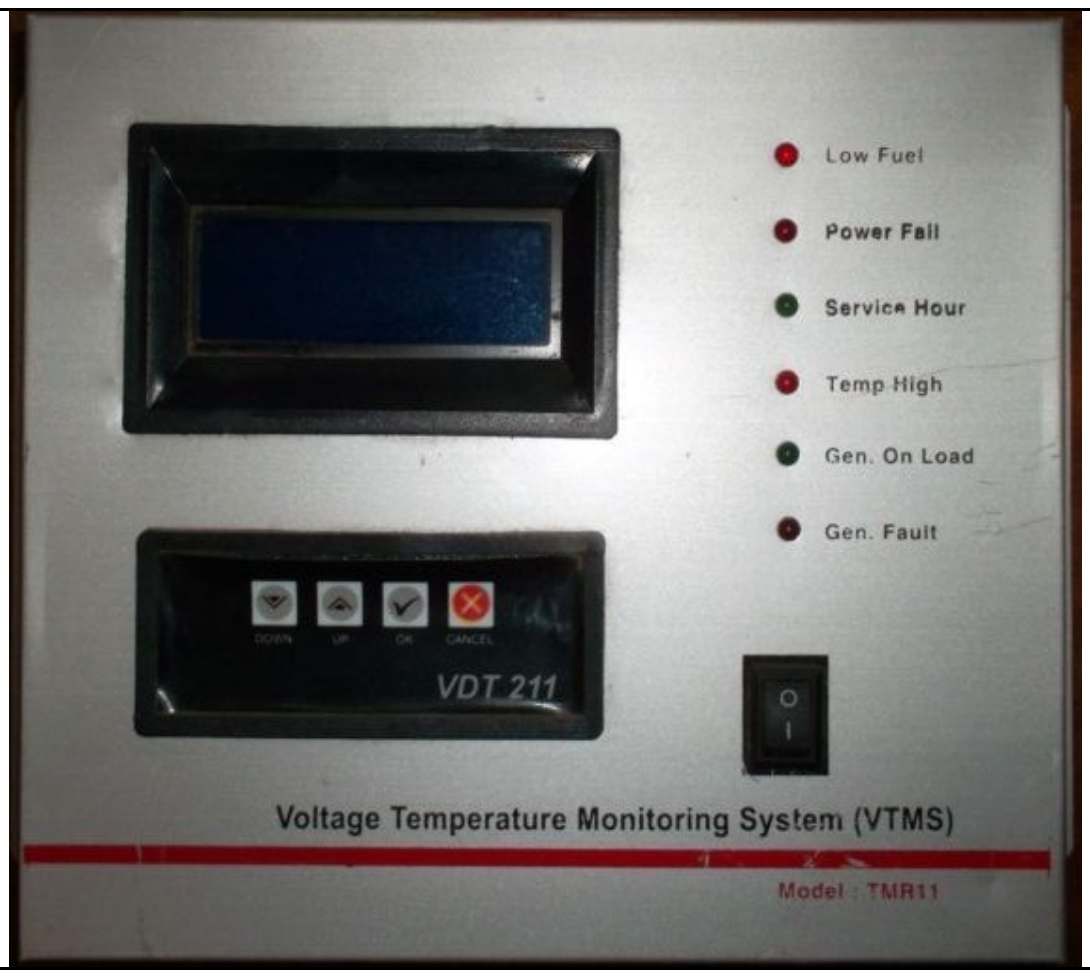

Figure 8(a). VTMS full product

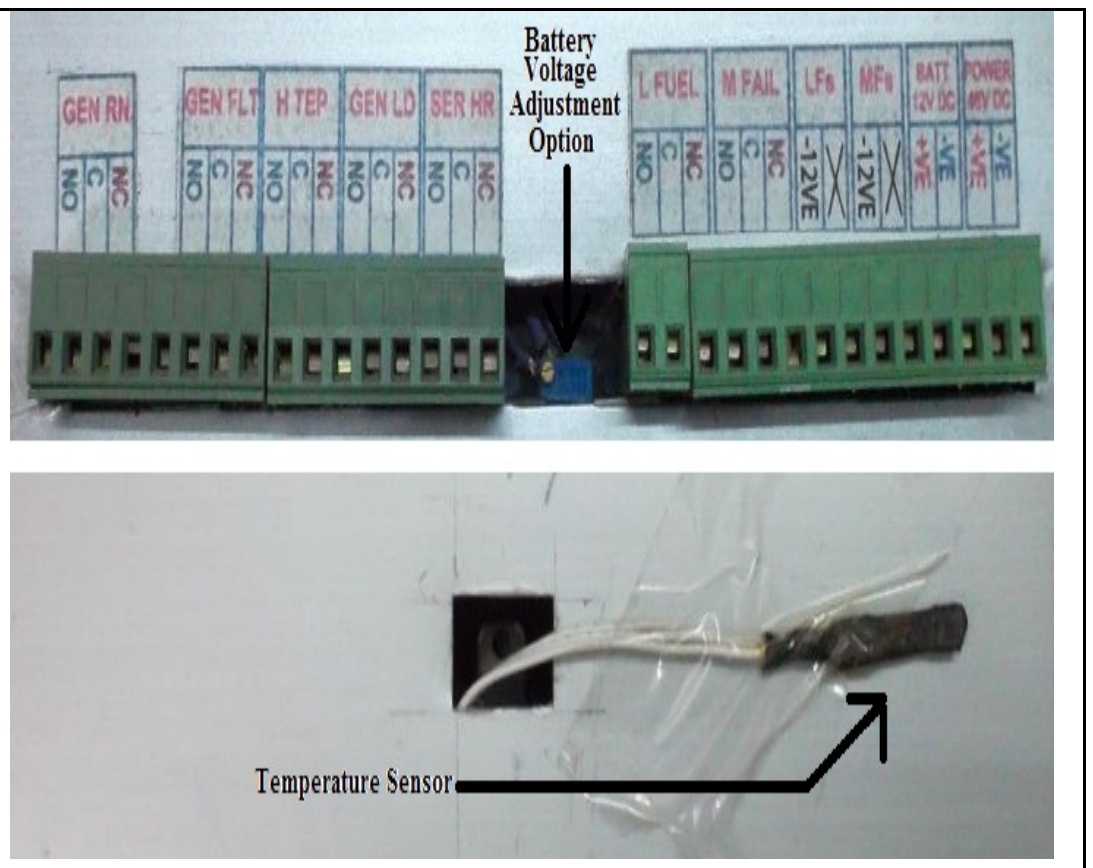

Figure 8(b). VTMS Connection Ports and Temperature Sensor

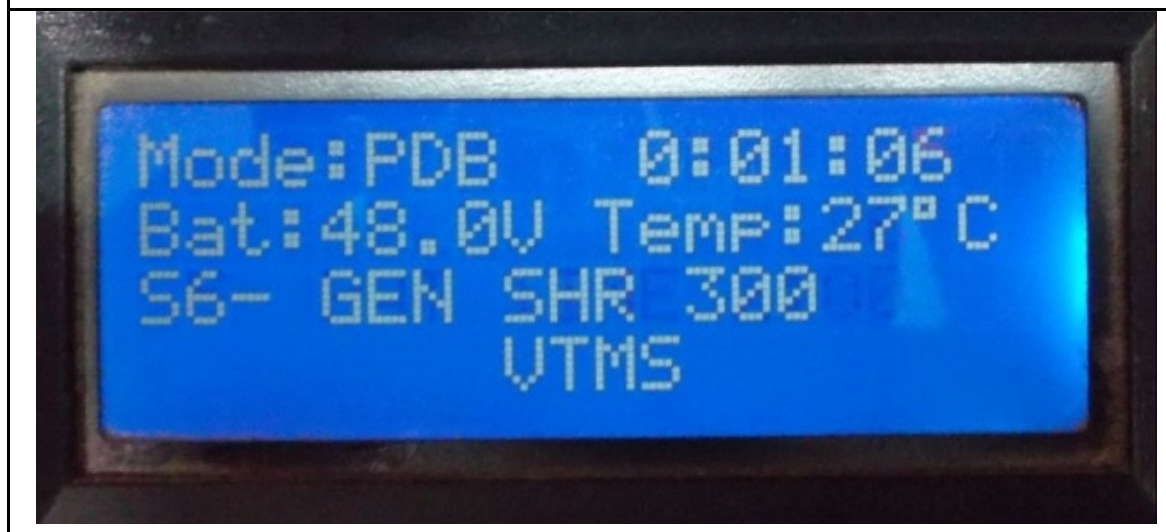

Figure 8(c). VTMS LCD display

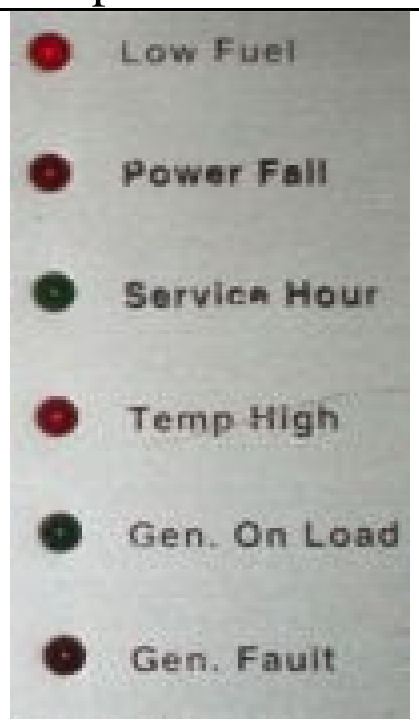

Figure 8(d). Fault indicator LED

# 5. CONCLUSION:

The reason and the actual motivation toward this work is to build a quality product which provides a full support to control the operation of the generator and bring synchronization between the battery and generator according to the PDB supply in BTS room. For the safety of the equipment of BTS room it is also necessary to control the temperature which is monitored by using VTMS. So it is a complete package for a BTS room to not only maintain the proper power system for BTS but also monitor the temperature for the safety of sophisticated equipments of BTS room. The manufacturing cost of VTMS is estimated to be 29 USD which is really cost effective.

## AUTHORS

**Sadeque Reza Khan** received B.Sc. degree in Electronics and Telecommunication Engineering from University of Liberal Arts Bangladesh and continuing his M.Tech in VLSI from National Institute of Technology Kernataka (NITK), India. Currently he is in study leave from his Institution where he was working as a lecturer in the department of Electrical and Electronic Engineering in Prime University, Bangladesh. His research interest includes VLSI, Microelectronics, Control System Designing and Embedded System Designing.

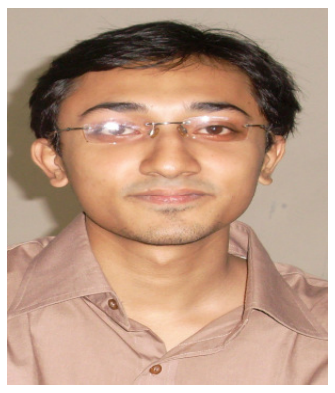

**Siddique Reza Khan** is a Computer Science Engineer from Military Institute of Science and Technology (MIST) under Dhaka University. His Research field covers Artificial Intelligence and Robotics. He also seeks some of his interest in control system designing.

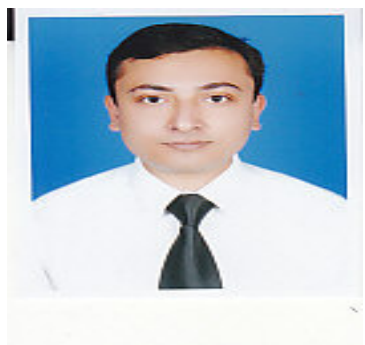

**Arifa Ferdousi** received B.Sc. and M.Sc. degree in ICE from University of Rajshahi, Bangladesh, in the year of 2007 and 2009 respectively. Currently she is working as a lecturer in the department of CSE in Varendra University, Rajshahi, Bangladesh. Her research interest includes electronics system designing, OFDM, Advanced LTE Wi-Max and Bangla speech recognition system using Neural Network. She is the member of Bangladesh Electronic Society (BES).

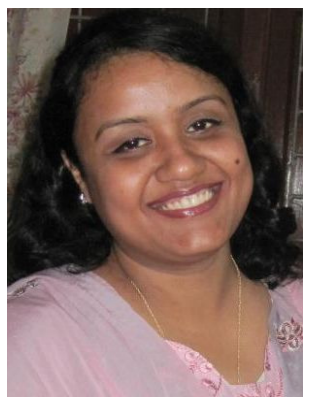